# The unusual X-ray and optical properties of the ultrasoft AGN Zwicky 159.034 (RE J 1237+264)

W.N. Brandt,[1] K.A. Pounds[2] and H. Fink[3]

[1] *Institute of Astronomy, Madingley Road, Cambridge CB3 0HA (Internet: niel@mail.ast.cam.ac.uk)*
[2] *X-ray Astronomy Group, Department of Physics & Astronomy, University of Leicester, University Road, Leicester LE1 7RH*
[3] *Max-Planck-Institut für Extraterrestrische Physik, D-85478, Garching, Germany*

**ABSTRACT**

Zwicky 159.034, one of the Seyfert galaxies identified with EUV sources detected during the *ROSAT* Wide Field Camera (WFC) all-sky survey, has unusual properties. The *ROSAT* Position Sensitive Proportional Counter (PSPC) 0.1–2.5 keV X-ray spectrum, obtained simultaneously with the WFC survey, appears extremely steep. Subsequent deeper pointed observations with *ROSAT* revealed that its 0.1–2.5 keV count rate had decreased by an extremely large amount (a factor of $\sim 70$). This is comparable to the variability amplitude seen in another ultrasoft Seyfert, E1615+061. There appears to be $\sim 10$ times as much flux in the soft component as in any expected hard tail suggesting that, in the absence of partial covering of the hard flux, *the soft flux cannot arise via reprocessing of the hard tail*. Its peculiar optical spectrum has permitted lines with widths $\sim 1\,200$–$1\,500$ km s$^{-1}$, and forbidden lines are weak or absent. Its optical spectrum shows evidence for variability, and might be used to study the role reprocessed X-rays play in the formation of optical lines.

## 1 INTRODUCTION

The majority of optical counterparts to the 383 EUV sources in the *ROSAT* WFC Bright Source Catalogue (BSC), as expected, are white dwarfs or active stars. A small group of extragalactic sources, all in directions of low interstellar hydrogen column density, include four Seyfert galaxies and three BL Lacertae objects. In general these galaxies were detected in the WFC survey because they were intrinsically bright soft X-ray/EUV sources and lie in low column directions.

Recent X-ray observations of Seyfert 1 galaxies have shown most of them to have underlying power-law continua with photon indices in the range $\Gamma = 1.9$–$2.0$ (e.g. Nandra & Pounds 1994), steepening to $\Gamma \sim 2.5$ below $\sim 1$ keV due to the presence of a 'soft excess' component. Three interesting exceptions are the BSC Seyferts RE J 1034+393, RE J 1442+352 (Mrk 478) and Zwicky 159.034 (RE J 1237+264, IC 3599). For these three objects, the ratio of soft (0.15–0.5 keV) to harder (1–2 keV) fluxes appears unusually high, being an order of magnitude larger than for a typical Seyfert galaxy. The first two objects are reported upon elsewhere (Gondhalekar et al. 1994, Pounds 1994, Puchnarewicz et al. 1995) and are both narrow-line Seyfert 1 galaxies (NLS1; for a succinct definition of this class of objects see Goodrich 1989). The third object, Zwicky 159.034, is the subject of this paper. It is a peculiar spiral galaxy in the region of the Coma cluster and is probably an outlying cluster member. Tifft & Gregory (1973) give a radial velocity corrected for local motions (assumed 300 km s$^{-1}$ rotation) of 6432 km s$^{-1}$ and a photographic magnitude of 15.7.

A value of the Hubble constant of $H_0 = 50$ km s$^{-1}$ Mpc$^{-1}$ and a cosmological deceleration parameter of $q_0 = \frac{1}{2}$ have been assumed throughout.

## 2 OBSERVATIONS AND DATA REDUCTION

*ROSAT* PSPC (Trümper 1983, Pfeffermann et al. 1987) observations were made of Zwicky 159.034 both during the all-sky survey and during the *ROSAT* pointed programme. The all-sky survey observation was made on 9 Dec 1990 and had a net exposure time of 777 s. Pointed observations include RP700552 (15 Dec 1991–2 Jun 1992, 4 827 s raw exposure in four observation blocks), RP701097 (30 Jun 1992–2 Jul 1992, 5 424 s raw exposure in three observation blocks), RP701098 (16 Jun 1992, 1 772 s raw exposure in one observation block), RP701099 (17 Jun 1992, 1 875 s raw exposure in one observation block) and RP701100 (18 Jun 1992, 2 011 s raw exposure in one observation block). The *ROSAT* pointed observations were performed in the standard 'wobble' mode to avoid accidental shadowing of sources by the coarse wire grid which forms part of the PSPC entrance window support structure.

Reduction and analysis of the PSPC data was performed with the Starlink ASTERIX X-ray data processing system and the XSPEC spectral fitting package (Shafer et al. 1991).

**Figure 1.** Positions relevant to this paper. The six pointed star is the *ROSAT* WFC all-sky survey position for Zwicky 159.034 and the five pointed star is its optical position. The five solid dots are the *ROSAT* X-ray centroid positions obtained during the pointed PSPC observations mentioned in the text. The open circle is the *ROSAT* PSPC all-sky survey position. The 'x' close to the X-ray and optical positions is the position for KN 16.14059 and the 'x' further away is the position for LB06690.

## 3 ANALYSIS

### 3.1 X-ray spatial analysis

X-ray imaging detectors such as the *ROSAT* PSPC improve the sensitivity and positional accuracy of cosmic X-ray observations relative to collimated detectors. Fig. 1 shows the sky near Zwicky 159.034. We have performed a careful positional analysis using optical, *ROSAT* PSPC and *ROSAT* WFC data to verify the positional coincidence and check for potential contaminating sources that might influence the results below. The optical position for Zwicky 159.034 is $\alpha_{2000} = $ 12h37m41.2s, $\delta_{2000} = $ 26d42m27.2s with an error of less than 1 arcsec (Klemola, Jones & Hanson 1987; McMahon & Irwin 1992). The *ROSAT* PSPC pointed X-ray centroid positions were obtained using the ASTERIX PSS software and have errors of less than 20 arcsec. The *ROSAT* PSPC all-sky survey X-ray centroid position has an error of less than 30 arcsec. The PSPC and optical positions agree very well. The *ROSAT* WFC survey position is $\alpha_{2000} = $ 12h37m35.9, $\delta_{2000} = $ 26d43m29.9s (Pye et al. 1995) and has a (statistical plus systematic) 95 per cent error circle of 94 arcsec. We believe the WFC source is Zwicky 159.034 despite the 101 arcsec positional offset, but it is worth noting that even if the WFC source does not correspond to the optical/X-ray source *this has no effect on the results of this paper*. We have searched the Palomar Optical Sky Survey blue and red plates and there is no evidence for any objects in addition to Zwicky 159.034 in the X-ray error circles. On the plates Zwicky 159.034 appears diffuse with an extent of $\sim$ 20 arcsec (Tifft & Gregory 1973). It has the bright central core expected for an active galaxy. There are no additional NED or SIMBAD sources within 5 arcmin.

The two other optical sources listed in Table 1 of Pounds et al. (1993) close to Zwicky 159.034 are also shown in Fig 1. KN 16.14059 is described as a 'nebula' in Wolf (1909) and Dixon & Sonneborn (1980). It is probably the same object as Zwicky 159.034 since there is no evidence for an additional nearby source on the Palomar plates. LB06690 is an unexceptional Luyten faint blue star with a photographic magnitude of 19.5 (Luyten 1962). The proper motion for this star does not appear to be in the literature so we were not able to correct its position for any potential proper motion. It lies 192 arcsec from the optical position of Zwicky 159.034.

Bade et al. (1995) have performed an entirely independent identification of the all-sky survey PSPC source and agree with our conclusions.

The 0.1–2.5 keV spatial profile of Zwicky 159.034 in all X-ray images is consistent with that of a point source convolved with the *ROSAT* XRT and PSPC spatial responses (Hasinger et al. 1992). In the temporal and spectral analyses below we have extracted the source counts for Zwicky 159.034 from circular source cells chosen to be large enough to ensure that all of the source counts are included, given the electronic 'ghost imaging' which widens the point spread function below $\sim$ 0.3 keV (Hasinger et al. 1992). Background counts were subtracted from the source cells using large circular source-free background cells. Corrections were included for PSPC dead time, vignetting and shadowing by the coarse mesh window support. We use these data in the temporal and spectral analyses below.

### 3.2 X-ray temporal analysis

Count rates during the pointed programme can only be reliably used for a representation of source fluxes as averages over the *ROSAT* wobble period of $\approx$ 400 s, as the path of the source in the detector should repeat with this period. We have used 400 s binning for all temporal analysis work. All of the pointed observations listed above have mean count rates in the range 0.05–0.08 count s$^{-1}$. There is a potential rise by a factor $\sim$ 2 in 1 700 seconds during the RP701100 pointing, but the low count rate and consequent large error bars make this variability of low statistical significance. Some variability may also exist in the second and third observation blocks of the RP700552 pointing. The all-sky survey observation shows marginal evidence for $\sim$ 20 per cent variability. The most striking variability, however, is apparent when the all-sky survey data is compared with the pointed phase data (shown in Fig. 2). During the survey Zwicky 159.034 was very bright and registered a mean count rate of 3.5 count s$^{-1}$, about 70 times more than the lowest count rate during the pointed observations. We will discuss this further below.

We have extensively searched for instrumental effects that might produce the variability seen in Zwicky 159.034, and can exclude them. We have compared the count rates of two other nearby sources (located at $\alpha_{2000} = $ 12h36m01s, $\delta_{2000} = $ 26d41m23s and $\alpha_{2000} = $ 12h36m57s, $\delta_{2000} = $ 26d57m03s) in the *ROSAT* pointed fields of view with their count rates during the all-sky survey, and they do not show correlated variability with Zwicky 159.034.

### 3.3 X-ray spectral analysis

Counts from the corrected circular source cells of observations were extracted into 256-channel, pulse-invariant spectra. We ignored channels 1–8 and rebinned the remaining channels so that there were at least 20 counts in each bin. Systematic errors of 2 per cent were added in quadrature to the data point rms errors, to account for residual uncertainties in the spectral calibration of the PSPC. We have used the standard survey response matrix for the all-sky survey analysis and the 1992 March response matrix (MPE No. 6) for the pointed analysis (this is the most appropriate matrix for the RP701097 data we present below). The expected systematic errors from these matrices are a few per cent, and are unlikely to alter the main results of our spectral fitting. We are keenly aware of remaining spectral calibration uncertainties with the PSPC, and refer the reader to the discussions of this issue in appendix A of Brinkmann et al. (1994) and appendix A and appendix B of Fiore et al. (1995).

**Figure 2.** Count rates during the all-sky survey and pointed *ROSAT* PSPC observations of Zwicky 159.034. The abscissa is Julian day minus 2 400 000 days.

**Figure 3.** Spectra for Zwicky 159.034 in the 0.1–2.5 keV band obtained during the all-sky survey (upper points) and RP701097 pointed phase observation (lower points). Note the statistics are much better for the all-sky survey observation even though it was only one seventh as long. A power-law and cold absorption fit to the all-sky survey data is shown with residuals. The same model is also shown fit to the pointed data below 0.4 keV, where only the power-law normalization has been allowed to vary. We include the two points above 0.4 keV in the plot for comparison purposes (see the text), although they were not used in the fit shown. Note that the spectrum remained steep after Zwicky 159.034 decreased in flux.

The Galactic neutral hydrogen column density in the direction of Zwicky 159.034 is $N_H = 1.25 \times 10^{20}$ cm$^{-2}$ (Stark et al. 1992). The errors for all fits are quoted for 68.3 per cent confidence (unless explicitly stated otherwise), taking all free parameters to be of interest other than absolute normalization. During the all-sky survey, the WFC registered $0.017 \pm 0.004$ count s$^{-1}$ in the S1a band ($\approx$100–170 eV), and the corresponding flux is consistent with the simultaneously measured PSPC survey flux. We have not used the WFC data in our analysis below as it does not provide significant additional constraints on the spectrum.

A single power-law and cold absorption fit to the all-sky survey data yields $\Gamma = 4.8^{+0.2}_{-0.2}$, $N_H = (5.2^{+0.6}_{-0.5}) \times 10^{20}$ cm$^{-2}$ and $\chi^2_\nu = 1.5$. Systematic residuals are visible throughout the spectrum, most notably at $\approx$0.15–0.45 keV and $\approx$0.8 keV. The derived spectrum is extremely steep, suggesting that Zwicky 159.034 is ultrasoft. The spectrum appears to remain steep above 0.5 keV. Fitting only the data above 0.5 keV gives $\Gamma = 4.7^{+0.3}_{-0.3}$. When we add simple blackbody and bremsstrahlung components to our models to accommodate the soft excess they do not improve the fits significantly ($\chi^2_\nu = 1.4$ for both soft excess models) and the power-law photon index becomes even larger ($\Gamma > 5$). Double blackbody and double bremsstrahlung soft component models also fail to improve the fits, as do lines and edges. The intrinsic $N_H$ obtained by our fits is somewhat surprising, and we also consider fits with $N_H$ fixed at the Galactic column. In this case we obtain a power-law photon index of $\Gamma = 3.06^{+0.03}_{-0.03}$ and $\chi^2_\nu = 6.4$. If we add a blackbody component to this model it becomes spectacularly powerful, completely dominating the spectrum below 1 keV. The obtained fit parameters are $\Gamma = 2.2^{+0.5}_{-0.6}$, $kT = 91^{+63}_{-29}$ eV and $A_{BB} = (6.3^{+0.6}_{-0.7}) \times 10^{-4}$ with $\chi^2_\nu = 2.2$. Note that the temperature of the soft excess is an extremely hot $\sim 10^6$ K, and that if this model can be regarded as having physical validity the soft excess would turn over within the *ROSAT* band. A similar temperature soft excess has been seen in the quasar E1346+266 (Puchnarewicz, Mason & Córdova 1994).

The pointed data have significantly poorer statistics than the all-sky survey data due to the lower count rate. We will present only the RP701097 data in detail since it is the longest exposure, was made over a relatively short span of time and showed no evidence for variability which might confuse the analysis. The other pointed data yield generally consistent results with the RE701097 data. A power-law and cold absorption fit is a satisfactory description of the data, giving $\Gamma = 4.6^{+2.0}_{-1.7}$, $N_H = (2.9^{+2.5}_{-2.0}) \times 10^{20}$ cm$^{-2}$ and $\chi^2_\nu = 0.9$. For comparison, a fit with $\Gamma$ fixed at 2 is unacceptable with $\chi^2_\nu = 3.6$ for 7 degrees of freedom (rejected at > 99 per cent confidence). Similarly, a fit with $\Gamma$ fixed at 2.5 has $\chi^2_\nu = 1.7$ for 7 degrees of freedom (rejected at 90 per cent confidence). Although the fit is poorly constrained, the fact that the best fit pointed and all-sky survey photon indices are consistent hints that the spectra may only differ in normalization. To test this possibility, we fit a power law with $\Gamma$ and $N_H$ set to the values derived from the all-sky survey simple power-law fit. This fit is formally unacceptable with $\chi^2_\nu = 2.8$, but when the two points above 0.4 keV are ignored the fit becomes acceptable with $\chi^2_\nu = 0.7$. The points above 0.4 keV fall below the fit to the data below 0.4 keV, perhaps hinting at a slight spectral steepening (see Fig. 3). However, the entire 0.1–2.5 keV pointed spectrum can be fit if $\Gamma$ and $N_H$ are allowed to vary within their 68 per cent confidence regions from the all-sky survey fit, so we are formally not able to quantify any potential small steepening between the all-sky survey and the pointed observation.

Since the spectrum for Zwicky 159.034 is very soft, changes in the intrinsic and Galactic neutral hydrogen column densities as a function of time might lead to significant changes in count rate. This possibility can be excluded from causing the large variability we see since the spectra in Fig. 3 do not dramatically change shape at low energies.

The flux we measure from Zwicky 159.034 is model dependent. If we use the power-law and cold absorption model from above, we find that during the survey the mean absorbed 0.1–2.5 keV flux was $(3.7^{+0.7}_{-0.6}) \times 10^{-11}$ erg cm$^{-2}$ s$^{-1}$ while the unabsorbed flux was $(1.2^{+0.8}_{-0.4}) \times 10^{-9}$ erg cm$^{-2}$ s$^{-1}$. This corresponds to a 0.1–2.5 keV isotropic luminosity of $(2.1^{+1.4}_{-0.6}) \times 10^{45}$ erg s$^{-1}$. It must be noted, however, that the high luminosity of this model depends sensitively on the potentially unphysical intrinsic cold absorption. If we use the power-law and blackbody model with $N_\mathrm{H}$ fixed at the Galactic column, we find that during the survey the mean absorbed 0.1–2.5 keV flux was $(3.7^{+0.7}_{-0.6}) \times 10^{-11}$ erg cm$^{-2}$ s$^{-1}$ while the unabsorbed flux was $(5.7^{+1.1}_{-1.0}) \times 10^{-11}$ erg cm$^{-2}$ s$^{-1}$. This corresponds to a 0.1–2.5 keV isotropic luminosity of $(1.1^{+0.2}_{-0.2}) \times 10^{44}$ erg s$^{-1}$. During RP701097, if the power law model is used the mean absorbed 0.1–2.5 keV flux was $(2.0^{+4.5}_{-1.3}) \times 10^{-13}$ erg cm$^{-2}$ s$^{-1}$ while the unabsorbed flux was $(3.1^{+20}_{-2.7}) \times 10^{-12}$ erg cm$^{-2}$ s$^{-1}$. This corresponds to a 0.1–2.5 keV isotropic luminosity of $(5.5^{+35}_{-4.9}) \times 10^{42}$ erg s$^{-1}$. If we compare the soft component fluxes with the hard tail fluxes (out to 200 keV) in our models above, we find that for all models considered there is always at least $\sim 10$ times more soft flux than hard flux.

It is important to note that the spectrum of Zwicky 159.034 is steep and remarkable *even when compared to* that of other *ROSAT* observed Seyfert 1 galaxies. Thus the general results of our spectral fitting are robust to potential future changes in the PSPC calibration.

### 3.4 Optical spectrum

Fig. 4 shows the optical spectrum of Zwicky 159.034. The strong lines show its active nature, but lines such as [O III] 4959 Å and [O III] 5007 Å are relatively weak. The spectrum reveals emission lines from H$\alpha$ 6563 Å, H$\beta$ 4861 Å, H$\delta$ 4101 Å, H$\gamma$ 4340 Å, He I 5876 Å, He I 7065 Å, He II 4686 Å, [O I] 8446 Å, [O III] 4959 Å, [O III] 5007 Å, [O I] 6364 Å and/or [Fe X] 6375 Å, [Fe XI] 7891 Å and [Fe XIV] 5303 Å. There is no evidence in the spectrum for emission from a potential foreground object.

Since the other two ultrasoft BSC Seyferts as well as many of the Puchnarewicz et al. (1992) ultrasoft AGN were narrow-line Seyfert 1 galaxies, we consider the possibility that Zwicky 159.034 is one as well. Goodrich (1989) has stated three criteria used to define the narrow-line Seyfert 1 class of objects, and Zwicky 159.034 satisfies all three criteria: (1) its Balmer lines are only slightly broader than its forbidden lines (the H$\alpha$ 6563 Å FWHM is 1 500 km s$^{-1}$ and the H$\beta$ 4861 Å FWHM is 1 200 km s$^{-1}$); (2) its ratio of [O III] 5007 Å to H$\beta$ is $< 3$; and (3) high ionization iron lines are present. Boller, Brandt & Fink (1995) have systematically studied the soft X-ray properties of 31 narrow-line Seyfert 1s and have discussed models for these objects. They find that narrow-line Seyfert 1s have generally steeper soft X-ray continua than normal Seyfert 1s (in agreement with the finding of Puchnarewicz et al. 1992), and this is consistent with what is seen in Zwicky 159.034. However, the relatively weak nature of the [O I] 6300 Å, [O I] 6364 Å, [O III] 4959 Å and [O III] 5007 Å lines, which are usually fairly strong in Seyfert 1s, suggests either an emaciated low density region or amplified emission from the high density region and optical continuum. The large change in X-rays may well be related to the peculiarity of the optical spectrum.

**Figure 4.** Observed optical spectrum of Zwicky 159.034 taken in May 1991 with the Faint Object Spectrograph on the Isaac Newton Telescope (Mason et al. 1995). Courtesy of Dr. M.G. Watson.

Tifft & Gregory (1973) describe a spectral observation of Zwicky 159.034 where the lines H$\beta$ 4861 Å, [O II] 3726 Å, [O III] 4959 Å, [O III] 5007 Å, Ca II 3934 Å and Ca II 3969 Å were noted (Ca II 3934 Å and Ca II 3969 Å lines were seen in absorption). We do not see evidence for [O II] 3726 Å, suggesting that the optical spectrum has changed. This might be expected in a galaxy where the X-ray flux can vary so dramatically.

## 4 DISCUSSION

Zwicky 159.034 showed a very large drop (a factor of $\sim 70$ in *ROSAT* PSPC count rate) in soft X-ray luminosity between 9 Dec 1990 and 15 Dec 1991 and has been relatively weak when studied since. It is not known if its drop was rapid or not. Contamination by a transient source is extremely unlikely due to the small error circles described above as well as the lack of evidence in the optical for any additional sources down to a Palomar plate magnitude limit of 20 in the blue and 21.5 in the red. The similarity of the shapes of the X-ray spectra between the all-sky survey and pointed observations is also suggestive of emission from a single source. Very large X-ray variability (a factor of $\sim 150$ in flux in the 0.1–0.28 keV band) was discovered by Piro et al. (1988) in another Seyfert 1 galaxy, E1615+061 (see also the optical follow-up work by Massaro, Nesci & Piro 1994). This object does not appear to be a narrow-line Seyfert 1 (Pravdo et al. 1981). One similarity between Zwicky 159.034 and E1615+061 is that both were ultrasoft when they were bright. However, the spectrum of E1615+061 became less soft when it became less luminous, in contrast to what has been seen so far in Zwicky 159.034 where the spectrum remains steep. The hardening seen by Piro et al. (1988) is consistent with the dramatic reduction of the intensity of a soft excess component or a change in its temperature which removes it from the instrument bandpass. Perhaps future X-ray monitoring of Zwicky 159.034 will show a similar spectral change. This might be analogous to the decay of some ultrasoft Galactic black hole candidates (BHC). While BHC eventually change to their hard states in decay, some remain ultrasoft for a significant time after they begin to dim (although if the decay time scales linearly with black hole mass the predicted decay time for Zwicky 159.034 is far longer than observed). It will be interesting to see if other objects that are optically similar to Zwicky 159.034 vary in X-rays by large amounts and, if so, over what timescales.

Two other objects, NGC 3628 and NGC 1068, also should be mentioned in connection with Zwicky 159.034. NGC 3628 is an interacting member of the Leo Triplet, and Dahlem et al. (1995) have discovered a drop by a factor of $\sim 20$ in its *ROSAT* band X-ray flux. Dahlem et al. (1995) suggest that NGC 3628 is either the most massive X-ray binary known so far, with a $> 75$ $M_\odot$ black hole, or an unusual low-luminosity AGN. It has recently been pointed out by Ulvestad, Antonucci & Goodrich (1995) that the archetypal Seyfert 2 galaxy NGC 1068 might well be a narrow-line Seyfert 1 if it were seen along the symmetry axis of its nucleus. It is interesting to note that in 1890 D.E. Packer reported an optical flare-up in NGC 1068 where its nucleus brightened to about 8th magnitude (de Vaucouleurs 1991).

When a simple power-law model is fit to the 0.1–2.5 keV spectrum of Zwicky 159.034, a photon index of $> 4$ is obtained. The soft flux measured from Zwicky 159.034 is at least $\sim 10$ times larger than that from any hard tail (barring dramatic high energy behavior). Thus, provided there is no strong obscuration which blocks the hard flux yet not the soft flux (e.g. the clouds of Guilbert & Rees 1988 and Nandra & George 1994), the soft flux *cannot* arise via reprocessing of the hard tail. In this case we would be seeing mostly intrinsic emission by matter, perhaps in an accretion disc or optically thin plasma, near the central black hole.

Another possibility worth consideration is that the observed large soft X-ray outburst was due to the tidal disruption of a star falling into or passing near the central black hole of Zwicky 159.034 (e.g. Rees 1990, Sembay & West 1993).

The large change in X-ray flux seen in Zwicky 159.034 can be used to explore the role reprocessed X-rays play in the formation of optical lines, and work on this matter is in progress.

## ACKNOWLEDGMENTS

We thank M.G. Watson for the optical spectrum. We thank Andy Fabian, Paul Nandra, Liz Puchnarewicz, Chris Reynolds and the members of the Cambridge X-ray seminar group for useful discussions. We thank Michael Dahlem, Harald Ebeling, Mike Irwin, Richard McMahon, Liz Puchnarewicz and Martin Ward for help with the Palomar sky survey plates and optical spectrum. We gratefully acknowledge help from members of the Institute of Astronomy X-ray group and financial support from the United States National Science Foundation and the British Overseas Research Studentship Programme (WNB). The *ROSAT* project is supported by the Bundesministerium für Forschung und Technologie (BMFT) and the Max-Planck Society.

## REFERENCES

Bade N., Fink H.H., Engels D., Voges W., Hagen H.-J., Wisotzki L., Reimers D., 1995, A&AS, in press
Boller Th., Brandt W.N., Fink H., 1995, A&A, *submitted*
Brinkmann W., et al., 1994, A&A, 288, 433
de Vaucouleurs G., 1991, The Observatory, 111, 122
Dahlem M., Heckman T.M., Fabbiano G., 1995, ApJ, in press
Dixon R.S., Sonneborn G., 1980, A Master List of Non-stellar Optical Astronomical Objects. Ohio State University Press, Columbus, Ohio
Fiore F., Elvis M., McDowell J.C., Siemiginowska A., Wilkes B.J., 1995, ApJ, *submitted*
Gondhalekar P.M., Kellett B.J., Pounds K.A., Matthews L., Quenby J.J., 1994, MNRAS, 268, 973
Goodrich R.W., 1989, ApJ, 342, 224
Guilbert P.W., Rees M.J., 1988, MNRAS, 233, 475
Hasinger G., Turner T.J., George I.M., Boese G., 1992, Legacy #2, The Journal of the High Energy Astrophysics Science Archive Research Center, NASA/GSFC
Klemola A.R., Jones B.F., Hansen R.B., 1987, AJ, 94, 501
Luyten W.J., 1962, A Search for Faint Blue Stars. University of Minnesota Press, Minneapolis, Minnesota
McMahon R.G., Irwin M.J., 1992, in MacGillivray H.T., Thomson E.B., eds, Digitised Optical Sky Surveys, Kluwer, Dordrecht, p. 417
Mason K.O., et al., 1995, MNRAS, *submitted*
Massaro E., Nesci R., Piro L., 1994, in Makino F., Ohashi T., eds, New Horizon of X-ray Astronomy: First Results from ASCA, Univ. Acad. Press, Tokyo, p. 397
Nandra K., George I.M., 1994, MNRAS, 267, 974
Nandra K., Pounds K.A., 1994, MNRAS, 268, 405
Pfeffermann E. et al., 1987, Proc. SPIE, 733, 519
Piro L., Massaro E., Perola G.C., Molteni D., 1988, ApJ, 325, L25
Pounds K.A., et al., 1993, MNRAS, 260, 77
Pounds K.A., 1994, in Makino F., Ohashi T., eds, New Horizon of X-ray Astronomy: First Results from ASCA, Univ. Acad. Press, Tokyo, p. 293
Pravdo S.H., Nugent J.J., Nousek J.A., Jensen K., Wilson A.S., Becker R.H., 1981, ApJ, 251, 501
Puchnarewicz E.M., Mason K.O., Córdova F.A., Kartje J., Branduardi-Raymont G., Mittaz J.P.D., Murdin P.G., Allington-Smith J., 1992, MNRAS, 256, 589
Puchnarewicz E.M., Mason K.O., Córdova F.A., 1994, MNRAS, 270, 663
Puchnarewicz E.M., Mason K.O., Siemiginowska A., Pounds K.A., 1995, MNRAS, *submitted*
Pye J., et al., 1995, MNRAS, *submitted*
Rees M.J., 1990, Science, 247, 817
Sembay S., West R.G., 1993, MNRAS, 262, 141
Shafer R.A., Haberl F., Arnaud K.A., Tennant A.F., 1991, XSPEC Users Guide. ESA Publications, Noordwijk
Stark A.A., Gammie C.F., Wilson R.W., Bally J., Linke R., Heiles C., Hurwitz M., 1992, ApJS, 79, 77
Tifft W.G., Gregory S.A., 1973, ApJ, 181, 15
Trümper J., 1983, Adv. Space Res., 4, 241
Ulvestad J.S., Antonucci R.R.J., Goodrich R.W., 1995, AJ, in press
Wolf M., 1909, Königstuhl Nebel-Liste. Veröffentlichung der Sternwarte zu Heidelberg